# Four annular structures in a protostellar disk less than 500,000 years old

Dominique M. Segura-Cox[1,2], Anika Schmiedeke[1], Jaime E. Pineda[1], Ian W. Stephens[3], Manuel Fernández-López[4], Leslie W. Looney[2], Paola Caselli[1], Zhi-Yun Li[5], Lee G. Mundy[6], Woojin Kwon[7,8], Robert J. Harris[2]

[1]Center for Astrochemical Studies, Max Planck Institute for Extraterrestrial Physics, Garching, 85748, Germany.
[2]Department of Astronomy, University of Illinois, Urbana, Illinois 61801, USA.
[3]Center for Astrophysics | Harvard & Smithsonian, 60 Garden Street, Cambridge, Massachusetts 02138, USA.
[4]Instituto Argentino de Radioastronomía (CCT-La Plata, CONICET; CICPBA), C.C. No. 5, 1894, Villa Elisa, Buenos Aires, Argentina.
[5]Astronomy Department, University of Virginia, Charlottesville, Virginia 22904, USA.
[6]Astronomy Department and Laboratory for Millimeter-wave Astronomy, University of Maryland, College Park, Maryland 20742, USA.
[7]Department of Earth Science Education, Seoul National University (SNU), 1 Gwanak-ro, Gwanak-gu, Seoul 08826, Republic of Korea.
[8]Korea Astronomy and Space Science Institute (KASI), 776 Daedeokdae-ro, Yuseong-gu, Daejeon 34055, Republic of Korea.

**Annular structures, or rings and gaps, in disks around pre-main sequence stars have been detected in abundance towards Class II objects ~1,000,000 years in age[1]. These structures are often interpreted as evidence of planet formation[1,2,3], with planet-mass bodies carving rings and gaps in the disk[4]. This implies that planet formation may already be underway in even younger disks in the Class I phase, when the protostar is still embedded in a larger scale dense envelope of gas and dust[5]. While younger disks likely play an important role in the onset of planet formation, only within the past decade have detailed properties of disks in the youngest star-forming phases begun to be observed[6,7]. Here we present 1.3 mm dust emission observations with 5 au resolution that show four annular substructures in the disk of the young (<500,000 years[8]) protostar IRS 63. IRS 63, a single Class I source located in the nearby Ophiuchus molecular cloud (at a distance of 144 pc[9]), is one of the brightest Class I protostars at (sub)millimeter wavelengths that also has a relatively large disk (>50 au)[10]. Multiple annular substructures observed towards disks at young times can act as an early foothold for dust grain growth, a prerequisite of planet formation. Whether planets already exist or not in the disk of IRS 63, it is clear that the planet formation process begins in the young protostellar phases, earlier than predicted by current planet formation theories[11].**

We studied the disk of IRS 63 with 5 au resolution via observing the emission from dust grains with the Atacama Large Millimeter/submillimeter Array (ALMA) at 1.3 mm. These observations revealed two concentric bright annular substructures (ring-like, R1 and R2) and two dark annular substructures (gap-like, G1 and G2), shown in Fig. 1. While dust structures within disks of Class I protostars have been previously observed[2,12-14], IRS 63 is the least evolved protostellar disk with multiple concentric dust annular substructures as indicated by different evolutionary indicators (see Methods). The annular substructures we observe towards the disk of the young protostar IRS 63 indicate that conditions for planetesimal formation likely begin from extremely early times, setting the stage for the first generation of planets to form.

The annular substructures in the disk of IRS 63 appear as plateaus of emission in the initial image (Fig. 2a) but are more apparent in the enhanced contrast image (Fig. 1; see Methods), which emphasizes compact structures and faint surface brightness enhancements. The annular substructures are also clearly revealed through the use of radiative transfer modeling. We created a smooth axisymmetric disk model heated by the central protostar and subtracted it from the data to create a residual image (Figs. 2b and 2c; see Methods). The residual image shows the same structure as the enhanced contrast image (Fig. 1) and reveals excess emission in the R1 and R2 rings, which is not accounted for by the smooth disk profile. The G1 and G2 gaps are also apparent. The residual emission in the north-west side of the disk is slightly brighter than in the south-east side, reflecting true asymmetry in the dust emission (see Methods and Extended Data Figs. 1, 2, 3, 4, 5, 6). It is possible that this disk asymmetry is is due to the formation of vortices, which are efficient at confining dusty material azimuthally and radially on timescales that can be <40,000 years[15].

We show the radial intensity profiles of the deprojected and azimuthally-averaged data, model, and residual images in Fig. 3a. To characterize the geometry of the rings we fit Gaussian profiles to the residual radial intensity profile (Fig. 3b; see Methods). The inner ring, R1, is located at 27 au with a width of 6 au. The outer ring, R2, is located at 51 au with a width of 13 au. Adding the Gaussian profiles of R1 and R2 together well reproduces the residual radial profile at radii > 20 au (see Fig. 3), and together with the underlying smooth disk model explains the majority of the disk emission in the vicinity of the rings. We use a similar method to measure the positions and widths of the gaps (Fig. 4). The inner gap, G1, is at a radius of 19 au with a width of 3.2 au. The outer gap at 37 au, G2, has a width of 4.5 au.

Gaps and rings have previously been detected towards disks, with the vast majority of ringed disks found around Class II pre-main sequence stars, a more-evolved phase where nearly all of the envelope has accreted or dissipated and the disk is the main component of circumstellar material[5]. To date, more than 35 of these Class II objects with mean ages of ~1,000,000 years[1] have been found with dust rings or gaps in their disks with ALMA[1,2,3]. The gaps are typically nearly devoid of dust emission at the (sub)millimeter wavelengths. The narrow and high contrast dust rings or gaps in these systems is often explained by the presence of planetary-mass bodies in the gaps, which shepherd the dust into sharply defined rings[4]. Indirect gas kinematic signatures of protoplanets in these disks also support this scenario[16]. These results indicate that planet formation must be both efficient and well underway already by the Class II phase. Recent dust mass measurements of Class II disks also indicate that observed dust depletion could be explained if significant mass is locked into planetesimals on timescales less than 0.1 - 1 Myr[17], placing early planet formation into the younger Class I phase and possibly even earlier. In comparison to the rings and gaps found in the older disks of Class II objects, the annular substructures in the younger disk of IRS 63 have dust emission even in the G1 and G2 gaps and are wider and lower contrast. In IRS 63, it is ambiguous if sizable planetary-mass bodies are creating these gaps.

If the gaps of IRS 63 are caused by planets that are already forming, the planet mass required to open each gap can be estimated from our dust observations. Hydrodynamic simulations for ranges of planet masses, disk parameters, and dust properties have allowed for the derivation of empirical relations connecting expected gap-opening planet masses, planet location, and gap width[18]. Assuming each gap is opened by one planet, for gap G1 the planet mass required is 0.47 $M_{Jupiter}$, and gap G2 requires a planet mass of 0.31 $M_{Jupiter}$ (see Methods). These gap-opening planet masses are high upper limits, because we determined the disk of IRS 63 to be more flared than the disks simulated to derive the empirical relation, and the estimated planet mass is sensitive to gap width while the G1 and G2 gaps are marginally resolved with our observations. Additionally, simulations show it is possible for a single planet to open multiple gaps[19]. The estimated planet masses in the young disk of IRS 63 are already comparable to Jupiter masses and are surprisingly large at these radii, considering that the early stage of planet formation faces serious barriers to grow to such large masses on short time scales as well as stopping runaway accretion at later stages.

While protostellar disks are certainly the sites of planet formation, in smooth disks the planet formation process faces a severe obstacle in the so-called "radial drift problem[20]." Dust in disks experience aerodynamic drag forces from the gas content of the disk, which causes a loss of angular momentum of the dust, and the dust particles will move inwards towards the protostar. This effect occurs on faster timescales for larger particles, and large solids will accrete onto the central protostar long before they can reach planetesimal sizes[20]. Annular substructures in planet-forming disks have been invoked to overcome the radial drift problem, and a more conservative interpretation of these substructures in the disk of IRS 63 is to consider that planets have not yet formed. Rings with higher densities of dust compared to the rest of the disk can occur where gas pressure maxima exist. The gas pressure maxima effectively trap the dust, preventing solids from further inspiraling towards the central protostar and creating a pile up of dust at a particular radius, forming a ring[21]. Further, it is possible these rings can achieve higher dust to gas ratios to act as zones of accelerated dust grain growth[22]. In IRS 63, we estimate an increase in the dust to gas ratio of 11% in R1 and 27% in R2 compared to other regions of the disk, though if the dust to gas ratio is constant throughout the disk, then the intensity variation might be due to variations in the dust absorption coefficient (see Methods). While observational evidence of grain growth has been detected towards young protostellar disks[23], until now multiple stable rings in which dust can reside and form larger solids in highly embedded disks have not been found. The rings observed in the disk around IRS 63 indeed demonstrate that this long-battled problem in planet formation can be overcome starting in the early stages of star formation.

Concentric dust annular substructures in disks are proposed to form via several mechanisms that do not require planets to be present, such as magneto-rotational instabilities[24], disk-winds in unevenly ionized disks[25], asymmetric accretion from the envelope to disk[26], or snowlines of volatile gas species freezing-out onto dust grains[27]. The temperature of the G2 gap is similar to the condensation temperature CO, and may be associated with the CO snowline (see Methods and Extended Data Fig. 7). We note that the G1, R1, and R2 substructures do not coincide with other major volatile species,[27] and if G2 formed due to a snowline, the other annular substructures likely formed via a different mechanism. Another ring formation mechanism which could be particularly applicable to the IRS 63 dust rings are self-induced dust traps[28]. These traps can form naturally from initially smooth disks on ~100,000 year timescales in simulations at relatively large radii when the aerodynamic drag, or back-reaction, of the dust on the gas is taken into account in addition to the drag of the gas on the dust, even in cases where the dust to gas ratio is 0.01[28]. For protostellar disk densities and dust grain sizes which are probed by millimeter observations, these often negligible back-reactions can become important and give rise to gas pressure maxima in the disk which traps particles. When combined with the effects of snowlines, back-reaction simulations have shown rings can form that are wider the further they are from the central protostar[29]. In the disk of IRS 63, the wider R2 ring is indeed at a larger radius, and the R1 and R2 rings may reflect a self-induced dust trap in the process of formation. While none of the annular substructure formation mechanisms can be completely confirmed or ruled out with the information presently at hand, the annular substructures of IRS 63 certainly present an observational benchmark against which simulations can be tested in an early evolutionary phase.

Dust within the outer dust disk (radii > 20 au), where the R1 and R2 rings are, could overcome the radial drift problem and eventually be incorporated into planets. The dust mass available to form future planets in this outer dust disk reservoir of material is 154 $M_{Earth}$ = 0.49 $M_{Jupiter}$, estimated by converting the continuum emission to mass[30] (see Methods). For a planetary embryo made of solids to trigger runaway accretion of gaseous material, which is required in the core accretion model of gas giant planet formation, the critical mass of 10 $M_{Earth}$ must be met by the solid core[31]. This condition is readily met in the young IRS 63 disk, even for a relatively low efficiency (< 10%[17]) of core formation out of the available dust grains. The large dust mass in the outer dust disk combined with the 27 and 51 au radii of the rings of IRS 63 is also consistent with evidence that Jupiter's core could have formed beyond 30 au in our own solar system and later migrated inward[32,33]. Giant planet cores may often form in the exterior regions of large disks starting from the early phases of star formation.

Even if planets have not yet formed or dust has not grown to large sizes yet, the dust rings in the IRS 63 disk at such a young evolutionary phase may serve as ideal zones for future dust grain growth and stable planetesimal formation. Even in the most conservative case of interpreting the annular substructures as variations in the disk density power law, rather than clear rings or gaps, these features are indications of dust beginning to pile up at particular radii in the disk. The structure of the disk likely has an impact on planet evolution starting early in the star formation process. Class I protostars remain embedded in a larger-scale envelope of gas and dust, which can replenish the disk as material is accreted, indicating that if planet formation is already beginning in the disk of IRS 63 then planets and protostars likely grow and evolve together from early times.

## References

1. Andrews, S. M. *et al.* The Disk Substructures at High Angular Resolution Project (DSHARP). I. Motivation, Sample, Calibration, and Overview. *Astrophys. J.* **869,** L41 (2018).
2. ALMA Partnership *et al.* The 2014 ALMA Long Baseline Campaign: First Results from High Angular Resolution Observations toward the HL Tau Region. *Astrophys. J.* **808,** L3 (2015).
3. van der Marel, N., Dong, R., di Francesco, J., Williams, J. P. & Tobin, J. Protoplanetary Disk Rings and Gaps across Ages and Luminosities. *Astrophys. J.* **872,** 112 (2019).
4. Huang, J. *et al.* The Disk Substructures at High Angular Resolution Project (DSHARP). II. Characteristics of Annular Substructures. *Astrophys. J.* **869,** L42 (2018).
5. André, P., Ward-Thompson, D. & Barsony, M. From prestellar cores to protostars: the initial conditions of star formation. In *Protostars and Planets IV* (eds Mannings, V., Boss, A. P. & Russell, S. S. ) 59–96 (Univ. Arizona Press, 2000).
6. Harsono, D. *et al.* Rotationally-supported disks around Class I sources in Taurus: disk formation constraints. *Astron. Astrophys.* **562,** 77 (2014).
7. Yen, H. *et al.* Signs of Early-stage Disk Growth Revealed with ALMA. *Astrophys. J.* **834,** 178 (2017).
8. Kristensen, L. E. & Dunham, M. M. Protostellar half-life: new methodology and estimates. *Astron. Astrophys.* **618,** A158 (2018).
9. Ortiz-León, G. N. *et al.* Gaia-DR2 Confirms VLBA Parallaxes in Ophiuchus, Serpens, and Aquila. *Astrophys. J. Lett.* **869,** L33 (2018).
10. Brinch, C. & Jørgensen, J. K. Interplay between chemistry and dynamics in embedded protostellar disks. *Astron. Astrophys.* **559,** A82 (2013).
11. Helled, R. *et al.* Giant Planet Formation, Evolution, and Internal Structure. In *Protostars and Planets IV* (eds Beuther, H., Klessen, R. S., Dullemond, C. P., & Henning, T. K.) 643-665 (Univ. Arizona Press, 2014).
12. Sheehan, P. D., & Eisner, J. A. Multiple Gaps in the Disk of the Class I Protostar GY 91. *Astrophys. J.* **857,** 18 (2018).
13. Sheehan, P. D., & Eisner, J. A. WL 17: A Young Embedded Transition Disk. *Astrophys. J.* **840,** L12 (2017).
14. de Valon, A. *et al.* ALMA reveals a large structured disk and nested rotating outflows in DG Tauri B. *Astron. Astrophys.* **634,** L12 (2020).
15. Bae, J., Hartmann, L., & Zhu, Z. Are Protoplanetary Disks Born with Vortices? Rossby Wave Instability Driven by Protostellar Infall. *Astrophys. J.* **805,** 15 (2015).
16. Pinte, C. *et al.* Nine localised deviations from Keplerian rotation in the DSHARP circumstellar disks: Kinematic evidence for protoplanets carving the gaps. *Astrophys. J. Lett.* **890,** L9 (2020).
17. Manara, C. F., Morbidelli, A. & Guillot, T. Why do protoplanetary disks appear not massive enough to form the known exoplanet population? *Astron. Astrophys.* **618,** L3 (2018).
18. Kanagawa, K. D. *et al.* Mass constraint for a planet in a protoplanetary disk from the gap width. *Publ. Astron. Soc. Jpn.* **68,** 43 (2016).
19. Dong, R., Li, S., Chiang, E., & Li, H. Multiple Disk Gaps and Rings Generated by a Single Super-Earth. II. Spacings, Depths, and Number of Gaps, with Application to Real Systems. *Astrophys. J.* **866,** 110 (2018).
20. Weidenschilling, S. J. Aerodynamics of solid bodies in the solar nebula. *Mon. Not. R. Astron. Soc.* **180,** 57-70 (1977).

21. Pinilla, P. *et al.* Trapping dust particles in the outer regions of protoplanetary disks. *Astron. Astrophys.* **538,** A114 (2012).
22. Youdin, A. N. & Goodman, J. Streaming Instabilities in Protoplanetary Disks. *Astrophys. J.* **620,** 459-469 (2005).
23. Harsono, D. *et al.* Evidence for the start of planet formation in a young circumstellar disk. *Nature As.* **2,** 646-651 (2018).
24. Gammie, C. F. Layered Accretion in T Tauri Disks. *Astrophys. J.* **457,** 355-362 (1996).
25. Suriano, S. S., Li, Z.-Y., Krasnopolsky, R., Suzuki, T. K. & Shang, H. The formation of rings and gaps in wind-launching non-ideal MHD discs: three-dimensional simulations. *Mon. Not. R. Astron. Soc.* **484,** 107-124 (2019).
26. Pineda, J. E. *et al.* A protostellar system fed by a streamer of 10,500 au length. *Nature As.* accepted (2020).
27. Zhang, K., Blake, G. A. & Bergin, E. A. Evidence of Fast Pebble Growth Near Condensation Fronts in the HL Tau Protoplanetary Disk. *Astrophys. J. Lett.* **806,** L7 (2015).
28. Gonzalez, J.-F., Laibe, G. & Maddison, S. T. Self-induced dust traps: overcoming planet formation barriers. *Mon. Not. R. Astron. Soc.* **467,** 1984-1996 (2017).
29. Vericel, A. & Gonzalez, J.-F. Self-induced dust traps around snow lines in protoplanetary discs. *Mon. Not. R. Astron. Soc.* **492,** 210-222 (2020).
30. Williams, J. P. *et al.* The Ophiuchus DIsk Survey Employing ALMA (ODISEA): Disk Dust Mass Distributions across Protostellar Evolutionary Classes. *Astrophys. J. Lett.* **875,** L9 (2019).
31. Pollack, J. B. *et al.* Formation of the Giant Planets by Concurrent Accretion of Solids and Gas. *Icarus* **124,** 62-85 (1996).
32. Öberg, K. I., & Wordsworth, R. Jupiter's Composition Suggests its Core Assembled Exterior to the $N_2$ Snowline. *Astrophys. J.* **158,** 194 (2019).
33. Bosman, A. D., Cridland, A. J., & Miguel, Y. Jupiter formed as a pebble pile around the $N_2$ ice line. *Astron. Astrophys.* **632,** L11 (2019).

**Acknowledgements**

ALMA is a partnership of ESO (representing its member states), NSF (USA) and NINS (Japan), together with NRC (Canada), MOST and ASIAA (Taiwan), and KASI (Republic of Korea), in cooperation with the Republic of Chile. The Joint ALMA Observatory is operated by ESO, AUI/NRAO and NAOJ. Authors DMSC, AS, JEP, and PC are grateful for support from the Max Planck Society. LWL acknowledges support from NSF AST-1910364. ZYL is supported in part by NASA 80NSSC18K1095 and NSF AST-1910106.

This research made use of Astropy, a community-developed core Python package for Astronomy[58].

**Author Contributions**

DMSC led the observing proposal and project, led the analysis, reduced the ALMA data, and wrote the manuscript. AS performed the radiative transfer modeling. JEP performed the MCMC modeling. IWS, MFL, LWL, ZYL, LGM, WK, and RJH contributed to the ALMA proposal. All authors discussed the results and implications, and commented on the manuscript.

**Competing financial interests**

The authors declare no competing financial interests.

**Corresponding author**

Correspondence and requests for materials should be addressed to Dominique M. Segura-Cox (dom@mpe.mpg.de).

## Figures

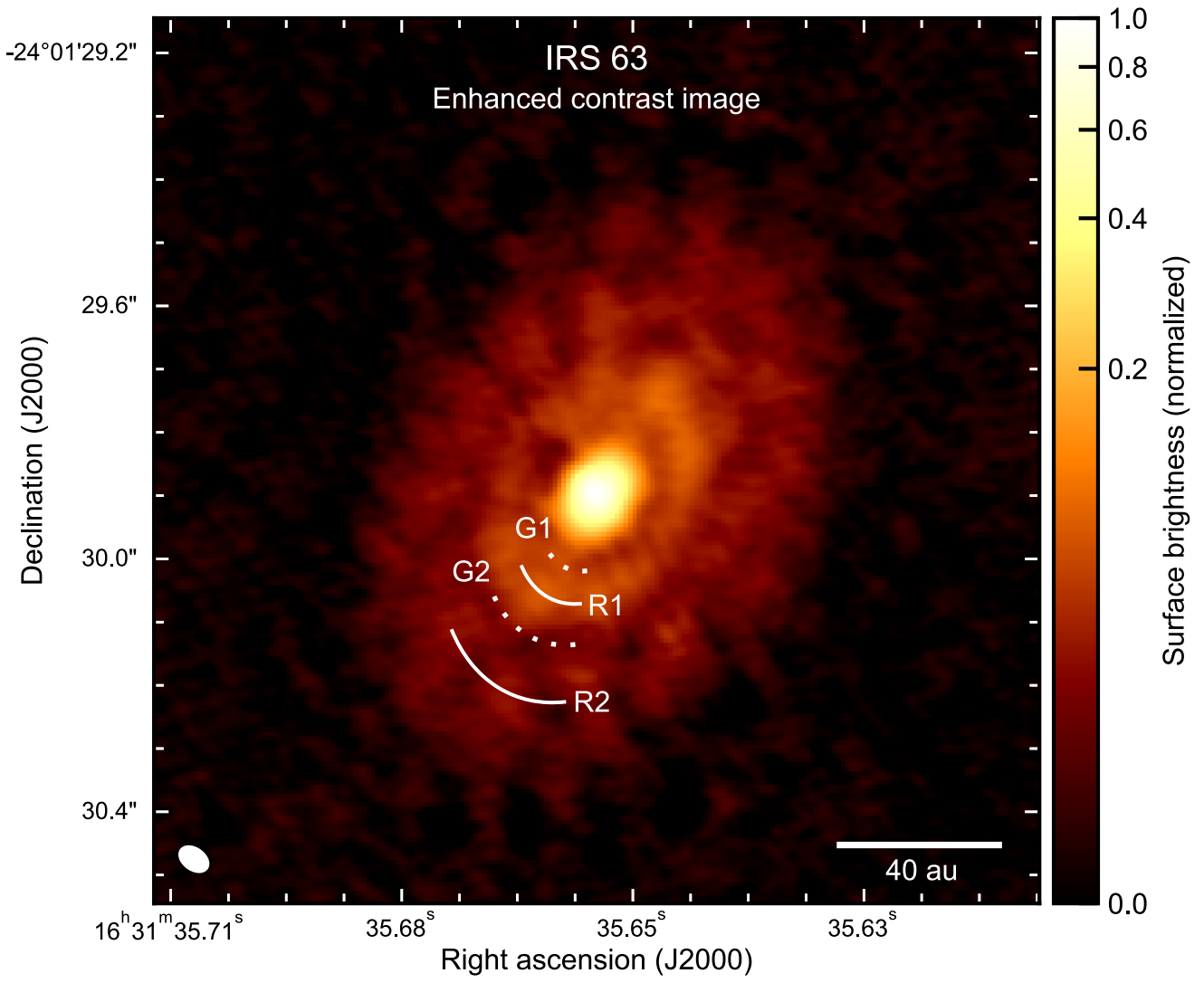

**Figure 1 | Image of the dust annular substructures around the Class I protostar IRS 63.** ALMA 1.3 mm dust continuum image, with an angular resolution of 0.05″ × 0.03″ (7 au × 4 au, represented by the white ellipse in the lower left). A scale bar is in the lower right. The color scale is the normalized surface brightness after a contrast enhancement technique is applied to the original data (see Methods). The enhanced contrast image clearly reveals two bright annular substructures (ring-like, R1 and R2, marked by solid white curves) and two dark annular substructures (gap-like, G1 and G2, marked by dotted white curves) in the disk surrounding the young protostar.

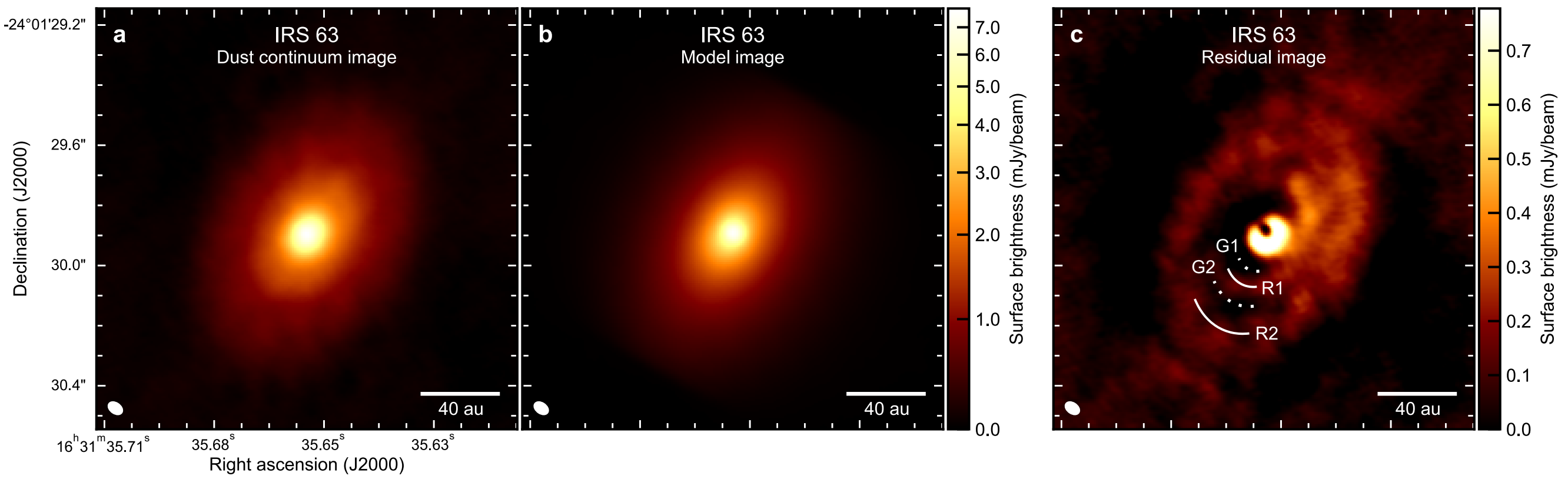

**Figure 2 | Radiative transfer modeling reveals dust annular substructures not explained by a smooth disk.** In each panel the resolution is shown by a white ellipse in the lower left, a scale bar is in the lower right, and the color scale represents the surface brightness. **a**, Original ALMA 1.3 mm dust continuum image, shown without contrast enhancement. **b**, Radiative transfer model of the IRS 63 disk, assuming a smooth geometry and imaged with the same resolution as the data (see Methods). **c**, The residual image of the observed data minus the model, yielding the R1, R2, G1, and G2 annular substructures consistent with those seen in Figure 1 (marked by white curves).

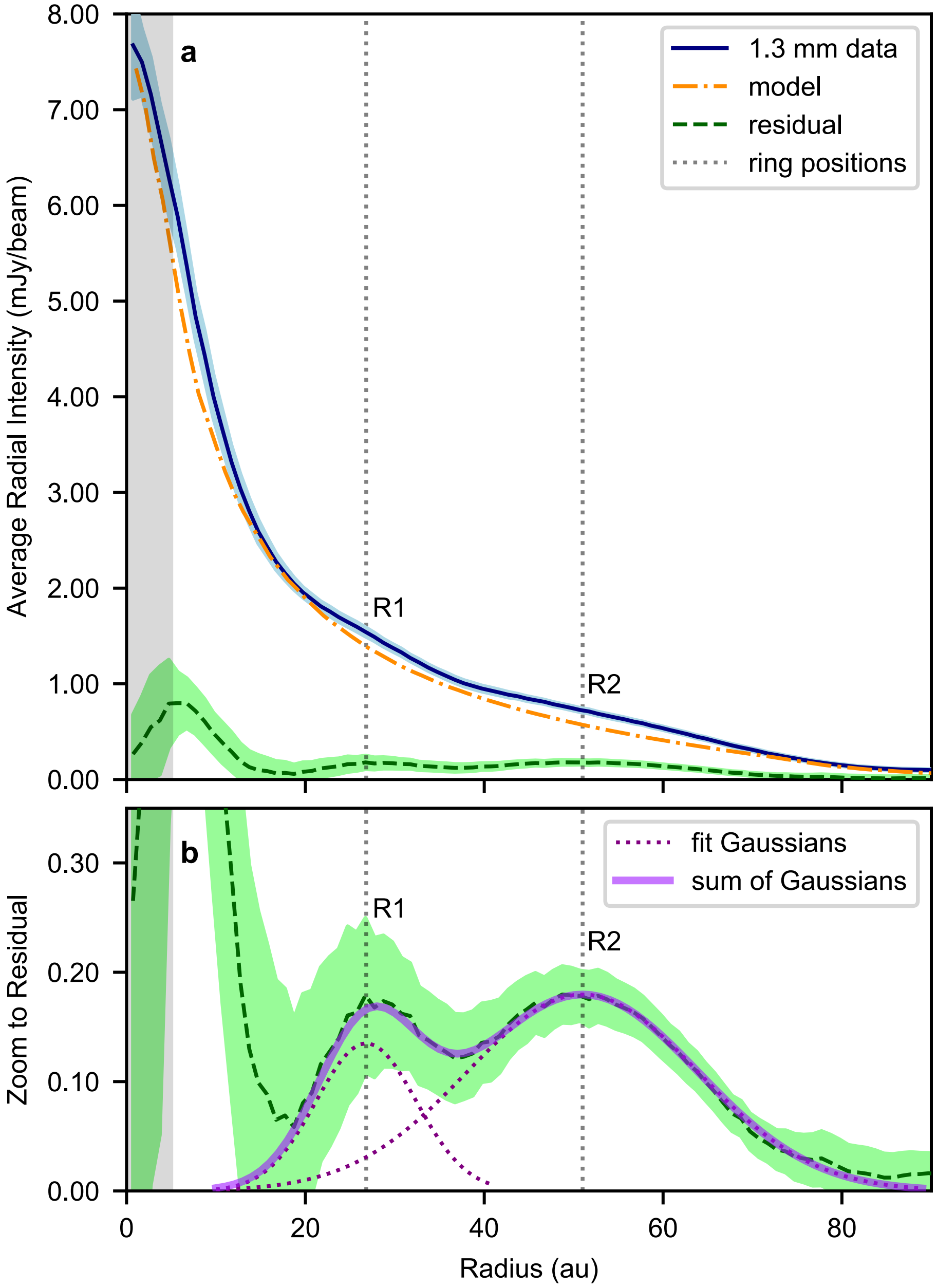

**Figure 3 | Ring positions and widths are measured from the radial profile of the residuals. a**, The radial intensity profiles of the images from the 1.3 mm data (solid blue line), the radiative transfer model (dot-dash orange line), and the data minus model residual (dashed green line) which have been deprojected, azimuthally-averaged, and radially binned into 1 au bins. The gray shaded area represents the resolution of the observations, and the vertical grey dotted lines show the positions of the R1 and R2 rings. The light blue and green ribbons represent the local standard deviation of each bin for the data and residual, respectively. **b**, A zoom-in of the y-axis to show the residual (dashed green line, representing extra emission in the rings not explained by a smooth disk profile), which was fit with two 1D Gaussians (dotted purple lines) to determine the positions and widths of the two rings. The sum of the fit ring Gaussians (solid thick purple line) well-reproduces the residual profile of disk exterior to 20 au.

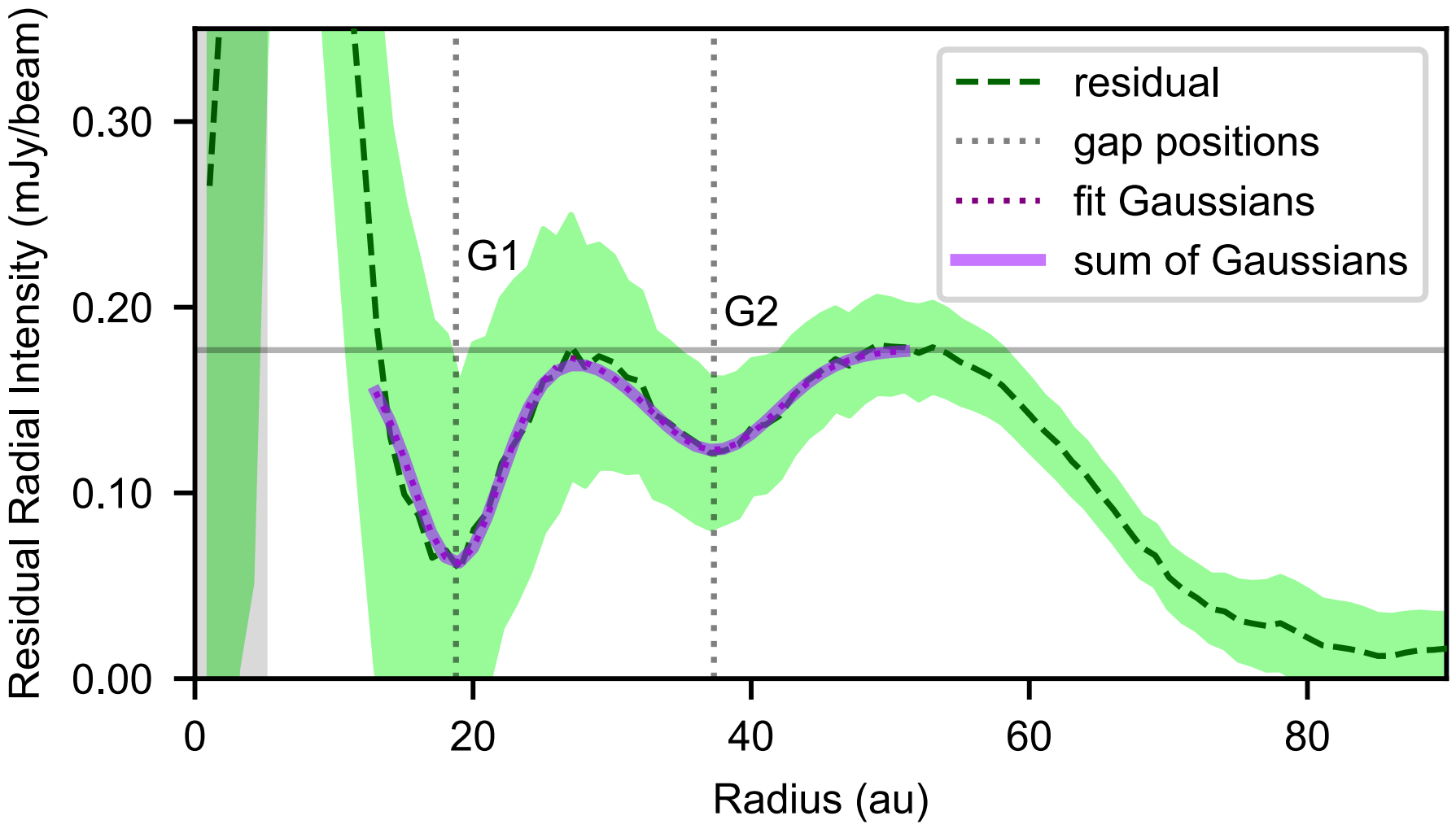

**Figure 4 | Gap positions and widths are measured from the radial profile of the residuals.** The same residual profile as seen in Figure 3 is again shown as a green dashed line, with the light green ribbon showing the local standard deviation of each bin. The gray shaded area represents the resolution of the observations, and the vertical grey dotted lines show the positions of the G1 and G2 gaps. The residual profile was fit with two 1D Gaussians (dotted purple lines), using the solid gray horizontal line as the baseline, to determine the positions and widths of the two gaps. The thick purple line is the sum of the fit gap Gaussians.

## Methods

### Observations and data reduction

We observed IRS 63 with ALMA under the Cycle 3 program 2015.1.01512.S. The observations were made on 2015 November 01 & 08, and 2017 September 18 in extended configurations with 41 m - 16.2 km baselines. Compact configurations with 15 m - 1.1 km baselines were made on 2016 May 24 and 2016 November 10 & 17. The total on-source integration time was 121 min. The correlator was configured for continuum observations with four 2 GHz spectral windows of 128 channels in dual polarization mode. The central frequency of the Band 6 observations was 233.002 GHz (1.3 mm) with a usable aggregate bandwidth of 7.5 GHz. Standard ALMA calibrators were used.

The extended and compact configuration data were first calibrated separately by ALMA staff using the standard pipeline calibration procedure with CASA[34] versions 4.7.2 and 4.7.0, respectively. We performed further manual flagging on the extended configuration observations to remove time periods of unstable phases and bad baselines and re-ran the pipeline with the additional flags. The estimated flux calibration uncertainty is 10%, and in this work we present statistical uncertainties unless otherwise noted. Self-calibration and imaging were performed with CASA 5.5.0. We concatenated all data and performed self-calibration iteratively, with shortest solution intervals of 10 s for phase and 60 s for amplitude cycles. The signal-to-noise ratio improved from 158 to 414 after self-calibration, with significant reduction of imaging artifacts.

The final continuum image (Fig. 2a) was created using multi-frequency synthesis, a Briggs weighting with robust parameter of 1.0, and multi-scale clean with scales of 0, 5, and 15 (0, 1, and 3× the synthesized beam). We chose the robust parameter of 1.0 as a compromise between spatial resolution and noise level in the final image. While the overall signal-to-noise ratio of the data is high, the substructures are low contrast compared to the overall disk emission. Lower robust parameter values, which would have resulted in smaller synthesized beam sizes, increased the noise level and made the detection of the annular substructures less clear. The synthesized beam is 0.05″ × 0.03″ (7 au × 4 au, assuming a distance of 144 pc[9]) with a position angle of 59.7°, and the image has a noise level of 18.4 μJy/beam. IRS 63 peaks at (α(2000),δ(2000)) = 16 h 31 m 35.6577 s, -24° 01′ 29.897″ with a peak flux of 7.78 mJy/beam and a total flux of 335.02 mJy.

To highlight the ringed structure in the disk, we created an enhanced contrast image (Fig. 1), which is a strategy used to bring out faint features in images that include bright components[35,36]. We smoothed the dust continuum image with a 2D Gaussian with a full-width at half-maximum of 0.06″, corresponding to 2× the minor axis of the synthesized beam, and multiplied all pixels in the smoothed image by a factor of 0.25. We subtracted the smoothed and scaled image from the original image and normalized the resultant image to produce the final enhanced contrast image. This procedure effectively removes the larger-scale emission to emphasize more compact structures in the disk, including the bright central area near the protostar and the more subtle rings in the outer dust disk.

### The youngest example of multiple annular substructures

The two previously youngest-known disks with multiple annular substructures are HL Tau[2] and GWYL 91[12]. HL Tau, with clear symmetric rings, is often regarded as lying between the Class I and Class II phases, while GWYL 91, with clumpy structures within rings, is firmly in the younger Class I phase. IRS 63 is even younger than these sources based on two evolutionary indicators: the bolometric temperature of an object ($T_{bol}$)[37] and the ratio of disk mass to envelope mass ($M_{disk}/M_{env}$)[38]. The values of both evolutionary indicators are expected to be lower in younger or less evolved sources with more surrounding envelope material. For IRS 63, the values of $T_{bol}$ and $M_{disk}/M_{env}$ are respectively 290 K[39] and 1.41[10]. The higher $T_{bol}$ of 576 K[40] and 370 K[39] and $M_{disk}/M_{env}$ of 9.43[12] and 8.26[12] for HL Tau and GWYL 91 respectively, demonstrate that IRS 63 is clearly the youngest amongst these three sources.

Models of protostars evolving with time from the onset of collapse show that for a 1.0 $M_{sun}$ protostar (appropriate for IRS 63, see below) with a $T_{bol}$ of 290 K, the age of the protostar is 130,000 - 140,000 years[41]. A lower protostellar mass would yield even lower ages for the same $T_{bol}$[41]. This value is also consistent with the Class I protostellar half-life of 135,000 years[8]. The determination of age from $T_{bol}$ is rather uncertain, yet a relative comparison between similar sources using these models identifies IRS 63 as the youngest protostellar disk with multiple annular substructures. The most conservative upper limit of the age of IRS 63 is ~500,000 years, the total lifetime of all embedded protostellar phases prior to the Class II pre-main-sequence phase[8,42].

Two other Class I protostars with dust structures in the disk are WL 17[13] and DG Tau B[14]. The disk of WL 17 has a hole in the inner 12 au of the disk and, with no other radial features in the dust, is considered an embedded transition disk[13] and is not a direct analogue to the multiple annular substructures of IRS 63. DG Tau B is a part of a wide binary system and has a dust spiral feature in addition to a broad dust ring[43]. The more complicated DG Tau B dust disk morphology likely reflects additional underlying physical processes at play compared to the isolated disk of IRS 63. L1527 IRS is an even younger Class 0/I protostar that was recently shown to have possible dust substructures in the disk[44]; however, due to the nearly edge-on disk orientation, the true morphology of the possible azimuthal substructures as rings or spirals is unclear.

**Radiative transfer disk model**

We produce a radiative transfer model of a smooth disk to compare to the original image and determine the geometry of the annular substructures. Because we use the model to determine the location and widths of the rings or gaps (see below), which will not vary with the relative intensity of the model, we choose to model a smooth disk which will best highlight the rings. Any annular substructures in the disk that are not explained by the smooth disk model are a result of disk radial density variations. We use the radiative transfer framework Pandora[45], which employs the 3D radiative transfer code RADMC-3D[46], version 0.41, for a self-consistent calculation of the dust temperature.

For the best fit model, we use CASA version 5.5.0, for post-processing of the synthetic model image. We sample the synthetic model image at the same uv-points as the original image with the CASA *ft* task and clean the resulting model uv-points with the same parameters as the original image. We subtract the model from the uv-data, and create the residual image with the same original imaging parameters. The resulting model image and residual images hence have the same beam as the original image (Fig. 2).

The density distribution $\rho(R, z, \phi)$ of the smooth disk[47] is given as

$$\rho(R, z, \phi) = \rho_0^{disk} \left(\frac{r_0}{R}\right)^{\beta - q} \exp\left[-\frac{1}{2}\left(\frac{z}{h(R)}\right)^2\right],$$

where $\rho_0^{disk}$ is defined by the disk mass $M^{disk}$, $r_0$ is the characteristic radius (set to 115.0 au), $\beta$ is the disk flaring power, $q$ is the surface density radial exponent ($\beta$ - $q$ is set to 2.25), and $h(R)$ is the disk scale height defined as

$$h(R) = h_0 \left(\frac{R}{r_0}\right)^{\beta}.$$

Here $h_0$ is the scale height at $r_0$. The disk starts at an inner radius $r_{in}$ (set to 1.0 au) and is truncated at an outer radius $r_{out}$. Heating is provided by a 1.0 $L_{Sun}$ point source located at the center of the disk[10]. A 1.0 $L_{Sun}$ protostar implies a mass of <1.0 $M_{Sun}$, since a portion of the luminosity is from accretion, and protostars are bigger and more luminous than main-sequence stars of the same masses. We also attempted modeling with a central luminosity value of 1.5 $L_{Sun}$, which always resulted in emission profiles too peaked towards the center, and the residuals were always lower for the 1.0 $L_{Sun}$ protostar family of models. We do not include heating by the interstellar radiation field, because the disk is heavily embedded in a thick envelope. The center of the disk was fixed to (α(2000),δ(2000)) = 16 h 31 m 35.6572 s, -24° 01′ 29.896″, determined from a Gaussian fit to the data. We set the viewing angles for ray tracing based on the position angle and inclination angle determined from the original image of the disk (see below).

We explore the parameter space $\left(\rho_0^{disk}, r_0, r_{out}, h_0, \beta - q\right)$ in a grid-based fashion and find the best fit for ($\rho_0^{disk}$ = $4 \times 10^9$ $H_2$ cm$^{-3}$, $r_0$ = 115 au , $r_{out}$ = 76 au, $h_0$ = 40 au, $\beta$ = 1.25, q = -1.0). We assume the dust to be covered in thin ice layers and coagulated at a density of $10^5$ cm$^{-3}$ [48]. At a wavelength of 1.3 mm, this yields a dust absorption coefficient of $\kappa$ = 0.7006 cm$^2$ g$^{-1}$. We also assume a canonical dust to gas ratio of $\epsilon_{model}$ = 0.01[49]. We explored values of $\beta$ between 1.0 and 1.5, encompassing both flatter and more flared disk profiles. We note that because this is a single-wavelength model, several degeneracies between parameters do exist. We give the parameter values of the best fit model for reproducibility, and we caution against taking any value as a concrete determination or constraint of that individual parameter. The overall model well reproduces the disk emission, and morphological deviations between the observations and model reflect physical disk structures not accounted for by the smooth disk model.

**Disk and annular substructure geometric properties**

Because the disk is well-resolved, a simple 2D Gaussian fit to the disk only fits the central bright peak near the protostar. To determine the radius and inclination of the dust disk, we draw a 10σ contour and measure the major and minor axis lengths ($b_{maj}$ and $b_{min}$, respectively) where they pass through the peak of the emission. This yields $b_{maj}$ = 1.13″ ± 0.02″ and $b_{min}$ = 0.79″ ± 0.02″, with the major axis having a position angle of 150° ± 2°. We define the radius as half the length of the major axis, the axis which is free from geometric projection effects, yielding a disk radius of 0.57″ or 82 au. The inclination angle with respect to the plane of the sky is then 45° ± 2°, defined as $\cos^{-1}(b_{min}/b_{maj})$.

To produce radial intensity profiles, we take into account the inclination angle of the disk, deproject the data assuming a circular disk geometry with the position angle determined above, and azimuthally average the data. We perform this for the original image, the model image, and the residual image. We then radially bin the data in 1 au bins each for the original, the model, and the residual images with the resulting averaged radial intensity profiles shown in Fig. 3.

We use the Python *SciPy curve_fit* routine[50], which employs $\chi^2$ minimization, to determine the annular substructure radii and widths from the radial profile of the residual image. For R1 and R2, we fit two 1D Gaussians to the residual radial profile. The two Gaussian profiles reproduce well the residual radial profile of the outer dust disk (exterior to 20 au), with R1 and R2 at radii of 27.0 ± 0.1 and 51.0 ± 0.1 au and respective Gaussian FWHM widths of 5.7 ± 0.1 au and 13.0 ± 0.1 au (Fig. 3). The uncertainties are the statistical uncertainties from the Gaussian fitting, and do not take into account uncertainties from the smooth disk model; however, the annular substructure locations and widths are not model dependent since these reflect variations in the observed radial intensity profile. For G1 and G2 we also fit two 1D Gaussians to the radial profile of the image with the same approach, taking the baseline for the Gaussians as the residual intensity value at 50 au. G1 and G2 are at radii of 19.0 ± 0.2 au and 37.0 ± 0.4 au with respective FWHM widths of 3.2 ± 0.2 au and 4.5 ± 0.4 au, where the uncertainties quoted are statistical (Fig. 4).

**Asymmetry in the disk**

Along with the clearly revealed annular substructures, the residual image of the best radiative transfer model also shows that the north-west side of the disk is slightly brighter than the south-east side of the disk (Fig. 2c). To determine the impact of the model centering on the morphology of the residual image, which may be a possible cause of asymmetry in residual images, we shifted the center of the best radiative transfer model image to (α(2000),δ(2000)) = 16 h 31 m 35.6567 s, -24° 01′ 29.890″. This position was found by minimizing the root mean square (*rms*) value of the difference between the data and model images for all possible model image centroid locations:

$$rms = \sqrt{\sum (I_{observed} - I_{model})^2 / N_{pix}},$$

where $I_{observed}$ and $I_{model}$ are the intensity at each pixel in the data and model images respectively, and $N_{pix}$ is the total number of pixels used to calculate the *rms* value. Compared to the original center of the best radiative transfer model, determined from a Gaussian fit to the data, the minimized center shows slightly lower amplitude residuals while the annular substructures and the disk asymmetry remains (Extended Data Fig. 1). These two positions are within a one pixel radius of each other, where one pixel is 0.007″ in size, and both positions are within a two pixel radius of the location of the peak intensity pixel. From this exercise, we rule out an off-center model image as the cause of the asymmetry.

The asymmetry does not occur along the minor axis of the disk, which would be expected if the asymmetry was a result of the flared disk geometry used in the radiative transfer model. To fully rule out the flared disk geometry as the cause of the disk asymmetry, we fit a smooth and geometrically flat analytic model profile to the data. The profile has been previously used to describe embedded protostellar disks[51], has a power law core with an exponential tail, and takes the form

$$\Sigma(R) = \Sigma_0 \left(\frac{R}{R_c}\right)^{-\gamma} \exp\left[-\left(\frac{R}{R_c}\right)^{2-\gamma}\right].$$

Here, Σ(R) is the intensity as a function of radius, $\Sigma_0$ is a normalization factor, $R_c$ is a characteristic radius, and $\gamma$ is the surface density gradient.

For the fitting, we used emcee[52], a Python Markov chain Monte Carlo (MCMC) sampler. The free parameters in the fit include $\Sigma_c$, $R_c$, $\gamma$, the coordinates of the centroid of the fit ($x_0$ and $y_0$), disk position angle (*PA*), and disk inclination angle (*IA*). We employed uniform priors for all parameters: 1.0 mJy/beam < $\Sigma_c$ < 2.0 mJy/beam, 10 au < $R_c$ < 100 au, 0.0 < $\gamma$ < 1.0, $x_0$ and $y_0$ within a 5 pixel radius of (α(2000),δ(2000)) = 16 h 31 m 35.6567 s, -24° 01′ 29.890″, 145.0° < PA < 155.0°, and 40.0° < IA < 50.0°. The run consisted of 3,840 walkers each with 5,000 steps and the first 200 steps were discarded as a burn-in phase. The resulting median values of the parameters and their uncertainties (see Extended Data Fig. 2 for a plot of the posterior distributions[53]) are $\Sigma_0 = 1.6985^{+0.0007}_{-0.0008}$ mJy/beam, $R_c = 51.96^{+0.02}_{-0.02}$ au, $\gamma = 0.4847^{+0.0002}_{-0.0002}$, $x_0 = -0.409^{+0.001}_{-0.001}$ pixel offset, $y_0 = -0.555^{+0.001}_{-0.001}$ pixel offset, $PA = 147.57^{+0.03}_{-0.03}$ °, and $IA = 44.95^{+0.02}_{-0.02}$ °. The uncertainties reflect the 16th and 84th percentiles of the samples from the marginalized distributions. The best fit $x_0$ and $y_0$ centroid of the emission corresponds to (α(2000),δ(2000)) = 16 h 31 m 35.6575 s, -24° 01′ 29.901″. Using the median values of the parameters from the MCMC modeling, we construct an analytic model image, and subtract it from the data image to produce a residual image (Extended Data Fig. 3). This method, with the flat disk geometry and fit centroid, also shows the four annular substructures and the asymmetry with excess emission in the north-west side of the disk. Therefore, we eliminate both model centering and flared disk geometry as being able to fully explain the asymmetric dust emission.

The asymmetry can also be seen as excess emission to the north-west side of the disk in the observed dust continuum image by taking radial intensity cuts passing through the disk centroid. We made three radial intensity cuts across the disk from edge to edge: one along the major axis with PA = 150.0°, one along the minor axis with PA = 60.0°, and one from the south-east to the north-west with PA = 105.0°. The radii are measured from the centroid outwards and deprojected assuming an IA of 45.0° and circular geometry. We performed this for three of the centroids discussed above: the peak intensity location, the Gaussian fit center, and the MCMC fit center (Extended Data Figs. 4, 5, 6 respectively). We then compare the intensities at each radius along a single cut on either side of the given center. In all centroid cases, the discrepancy between the two sides of the cut is greater for the south-east to north-west cut compared to the major and minor axis cuts. The north-west side of the disk has higher intensity values at most radii compared to the south-east side of the disk, and the slight asymmetry is hence also detectable without the use of radiative transfer or analytic disk models.

**Upper limit on planet mass from gap width**

If we assume that the origin of the dust annular substructures in the disk of IRS 63 arise from planets clearing disk material, and that the dust is well coupled to the gas content of the disk, we can apply the empirically derived relationship between gap width and planet mass[18]. The relationship, derived from 2D hydrodynamic simulations for a variety of planet masses and disk parameters[18] is

$$\frac{\Delta_{gap}}{R_p} = 0.41 \left(\frac{M_p}{M_*}\right)^{1/2} \left(\frac{h_p}{R_p}\right)^{-3/4} \alpha_{visc}^{-1/4},$$

where $\Delta_{gap}$ is the gap width, $R_p$ is radial location of the planet, $M_p$ is the mass of the planet, $M_*$ is the mass of the central star, $h_p$ is the scale height of the disk at $R_p$, and $\alpha_{visc}$ is the viscous parameter[54]. We estimate $\Delta_{gap}$ and $R_p$ from our measured Gaussian FWHM gap widths and gap locations. We take $M_*$ to be 0.8 $M_{Sun}$[10], calculate $h_p$ at $R_p$ based on the scale height prescription we used for the radiative transfer model, and set $\alpha_{visc} = 10^{-3}$.

We assume each gap is opened by a single planet lying at each gap's location. If the gaps are not fully resolved by our observations and are much narrower than the minor axis of the synthesized beam, this method will overestimate $M_p$. Since for both gaps $\Delta_{gap}$ is marginally resolved, this translates to an extreme upper limit for the $M_p$ required to open each gap. For gap G1 we find a highest upper limit of $M_p = 0.47 \pm 0.06$ $M_{Jupiter}$ and for G2 we find a highest upper limit of $M_p = 0.31 \pm 0.06$ $M_{Jupiter}$, where the uncertainties arise from the Gaussian fits of gap locations and widths. We point out that the true uncertainties are likely much larger since we found the disk of IRS 63 to be more flared than the simulated disks[18] used to derive the empirical relationship. Additional uncertainties arise because $M_p$ is sensitive to gap width and $M_*$, which is relatively unconstrained[10].

**Dust to gas ratios in the rings**

In our radiative transfer model, we hold the dust to gas ratio ($\epsilon_{model}$) constant. The extra emission in the rings could be explained by higher concentrations of dust in these rings relative to the other regions of the disk. This would be reflected by an increased dust to gas ratio since intensity is proportional to the dust to gas ratio, $I \propto \epsilon$. To quantify the increased dust to gas ratio needed to explain the residual emission at a particular radius, $\epsilon_{observed}(R)$, assuming all other parameters do not vary we divide the observed intensity, $I_{observed}(R)$, by the model intensity, $I_{model}(R)$, yielding the relation

$$\frac{I_{observed}(R)}{I_{model}(R)} = \frac{\epsilon_{observed}(R)}{\epsilon_{model}}.$$

In ring R1 at 27 au $\epsilon_{observed} = 0.0111 \pm 0.0004$, and in ring R2 at 51 au $\epsilon_{observed} = 0.0127 \pm 0.0005$. The excess ring emission could instead be similarly derived as increased dust absorption coefficients of $\kappa = 0.78 \pm 0.03$ cm$^2$ g$^{-1}$ in R1 and $\kappa = 0.89 \pm 0.03$ cm$^2$ g$^{-1}$ in R2. While these values of $\epsilon_{observed}$ do not yet reach the order of unity corresponding to the most efficient avenue of solid growth via the streaming instability[22], further condensing of dust in the rings may trigger rapid growth at later times.

**Correspondence between annular substructure locations and snowlines**

To determine the in-band spectral index ($\alpha$) and the dust temperature (T), we follow the methodology of a previous study of a snowline in a Class I disk[55]. We split the data into two separate datasets (with frequencies of $\nu_1$ = 241.002 GHz and $\nu_2$ = 225.002 GHz, with respective fluxes of $F_1$ and $F_2$) and calculate $\alpha = \ln(F_1/F_2) / \ln(\nu_1/\nu_2)$. We produced the radial profile of $\alpha$ with 1 au bins (Extended Data Fig. 7a). We calculated the optical depth, $\tau$, via graybody fitting[55] of the data at both frequencies, with the resulting $\tau$ radial profile shown in Extended Data Fig. 7b. Where $\tau > 3$, the graybody fit converged to the brightness temperature of the observations. Because of degeneracies in the optically thin regime between $T$ and $\tau$, where $\tau < 3$ we extrapolate $T$ to larger radii with a $T \propto r^{-1/2}$ power law (Extended Data Fig. 7c).

Using the observationally-derived $T$ of the disk, we compare the locations of the annular structures with the condensation temperatures of major volatile gas species, which have been previously computed for gas densities of $10^{10} - 10^{13}$ $H_2$ $cm^{-3}$[27]. G2 corresponds with the location of the CO snowline[27]. The spectral index increases across G2 while the optical depth decreases sharply, in agreement with previous observations and model signatures of snowlines[55] (Fig. 7). Further, we find $\tau > 1$ through most of the disk, indicating that self-scattering may be important and the true radial variations in intensity may have higher contrast than we observe here[56]. Self-scattering can also explain the anomalously low $\alpha < 2$ at the disk center[57], where $\alpha = 2$ is the blackbody Rayleigh-Jeans limit.

**Estimation of dust mass**

To estimate the dust mass of the disk, we follow the canonical flux to mass conversion used to estimate disk mass of IRS 63 from continuum at the same wavelength in a previous study[30], and we maintain the same assumptions as the previous study including $T$ = 20 K. This method assumes optically thin emission with a uniform temperature in the disk. The dust mass in the outer dust disk (at radii > 20 au, corresponding to the inner edge of the inner dust ring determined from the 1D Gaussian fitting of the radial profile) is 154.31 $M_{Earth}$ = 0.49 $M_{Jupiter}$, with a 3σ mass sensitivity of 0.03 $M_{Earth}$. If we instead adopt a dust temperature of $T$ = 40 K, which is high for most of the disk at radii > 20 au, this yields a more conservative dust mass of 66.44 $M_{Earth}$ = 0.21 $M_{Jupiter}$, with a 3σ mass sensitivity of 0.01 $M_{Earth}$. Even considering the more conservative dust mass estimate and taking into account low mass accumulation efficiency, there is enough dust mass in the outer dust disk where the rings lie for the planetesimal core accretion model to proceed[31].

**Data availability**

This paper makes use of ALMA data with the project code: ADS/JAO.ALMA#2015.1.01512.S. The archival data are available at http://almascience.eso.org/aq/ by querying the project code. The final calibrated version of the data analyzed in this work are available from the Harvard Dataverse with the identifier doi:10.7910/DVN/LPVDSF. Other material in this work is available from the corresponding author on reasonable request.

**Code availability**

The radiative transfer modeling makes use of RADMC-3D, which is is publicly available at http://www.ita.uni-heidelberg.de/~dullemond/software/radmc-3d/. The radiative transfer modeling also uses the Pandora framework for the setup of and interfacing with RADMC-3D; Pandora is not open source and is available upon request. The MCMC modeling uses emcee, which is publicly available at https://emcee.readthedocs.io/en/stable/.

**References**

34. McMullin, J.P., Waters, B., Schiebel, D., Young, W. & Golap, K. CASA Architecture and Applications. In *Astronomical Data Analysis Software and Systems XVI* (eds Shaw, R. A., Hill, F. & Bell, D. J.) 127–130 (Astron. Soc. Pacif. Conf. Ser. Vol. 376, 2007).
35. Pérez, L. M. *et al.* Spiral density waves in a young protoplanetary disk. *Science* **353,** 1519-1521 (2016).
36. van der Marel, N., Williams, J. P. & Bruderer, S. Rings and Gaps in Protoplanetary Disks: Planets or Snowlines? *Astrophys. J. Lett.* **867,** L14 (2018).
37. Myers, P. C. & Ladd, E. F. Bolometric Temperatures of Young Stellar Objects. *Astrophys. J. Lett.* **413,** L47 (1993).
38. Crapsi, A., van Dishoeck, E. F., Hogerheijde, M. R., Pontoppidan, K. M. & Dullemond, C. P. Characterizing the nature of embedded young stellar objects through silicate, ice and millimeter observations. *Astron. Astrophys.* **486,** 245 (2008).
39. Enoch, M. L., Evans, N. J., II, Sargent, A. I. & Glenn, J. Properties of the Youngest Protostars in Perseus, Serpens, and Ophiuchus. *Astrophys. J.* **692,** 973-997 (2009).
40. White, R. J., & Hillenbrand, L. A. On the Evolutionary Status of Class I Stars and Herbig-Haro Energy Sources in Taurus-Auriga. *Astrophys. J.* **616,** 998-1032 (2004).
41. Young, C. H. & Evans, N. J., II. Evolutionary Signatures in the Formation of Low-Mass Protostars. *Astrophys. J.* **627,** 293-309 (2005).
42. Evans, N. J., II *et al*. The Spitzer c2d Legacy Results: Star-Formation Rates and Efficiencies; Evolution and Lifetimes. *Astrophys. J. Supp.* **181,** 321-350 (2009).
43. Garufi, A. *et al.* ALMA chemical survey of disk-outflow sources in Taurus (ALMA-DOT): I. CO, CS, CN, and H2CO around DG Tau B. *Astron. Astrophys.* **636,** A65 (2020).
44. Nakatani, R. *et al.* Substructure Formation in a Protostellar Disk of L1527 IRS. *Astrophys. J. Lett.* **895,** L2 (2020).
45. Schmiedeke, A. *et al.* The physical and chemical structure of Sagittarius B2. I. Three-dimensional thermal dust and free-free continuum modeling on 100 au to 45 pc scales. *Astron. Astrophys.* **588,** A143 (2016).

46. Dullemond, C. P. *et al.* RADMC-3D: A multi-purpose radiative transfer tool. *Astrophysics Source Code Library*. Record ASCL:1202.015 (2012).
47. Pineda, J. E. *et al.* The Enigmatic Core L1451-mm: A First Hydrostatic Core? Or a Hidden VeLLO? *Astrophys. J.* **743,** 201 (2011).
48. Ossenkopf, V. & Henning, Th. Dust opacities for protostellar cores. *Astron. Astrophys.* **291,** 943-959 (1994).
49. Hildebrand, R. H. The determination of cloud masses and dust characteristics from submillimetre thermal emission. *Q. J. R. Astron. Soc.* **24,** 267–282 (1983).
50. Virtanen, P. *et al.* SciPy 1.0--Fundamental Algorithms for Scientific Computing in Python. Preprint at http://ArXiv.org/abs/1907.10121 (2019).
51. Andrews, S. M., Wilner, D. J., Hughes, A. M., Qi, C. & Dullemond, C. P. Protoplanetary Disk Structures in Ophiuchus. *Astrophys. J.* **700,** 1502-1523 (2009).
52. Foreman-Mackey, D., Hogg, D. W., Lang, D. & Goodman, J. emcee: The MCMC Hammer. *Publ. Ast. Soc. Pacific* ***1*25,** 306 (2013).
53. Foreman-Mackey, D. corner.py: Scatterplot matrices in Python. *J. Open Source Softw.* **1,** 24 (2016).
54. Shakura, N. I., & Sunyaev, R. A. Reprint of 1973A&A....24..337S. Black holes in binary systems. Observational appearance. *Astron. Astrophys.* **500,** 33-51 (2009).
55. Cieza, L. A. *et al.* Imaging the water snow-line during a protostellar outburst. *Nature*. **535,** 258-261 (2016).
56. Zhu, Z. *et al.* One Solution to the Mass Budget Problem for Planet Formation: Optically Thick Disks with Dust Scattering. *Astrophys. J. Lett.* **877,** L18 (2019).
57. Liu, H. B. The Anomalously Low (Sub)Millimeter Spectral Indices of Some Protoplanetary Disks May Be Explained By Dust Self-scattering. *Astrophys. J. Lett.* **877,** L22 (2019).
58. Astropy Collaboration *et al.* Astropy: A community Python package for astronomy. *Astron. Astrophys.* **558,** A33 (2013).

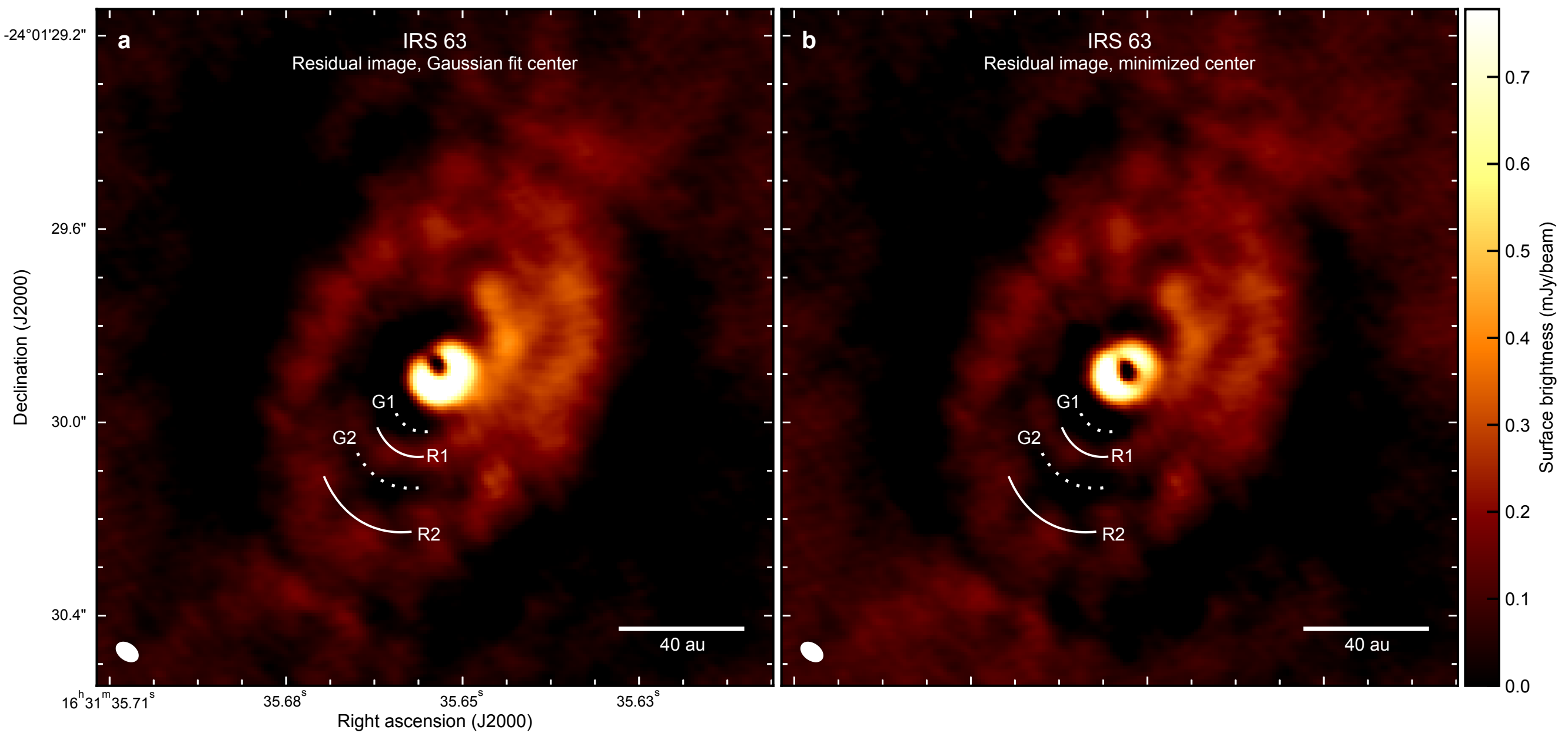

**Extended Data Figure 1 | Asymmetry remains after re-centering the best radiative transfer model. a**, The residual of the data minus the best radiative transfer model, which had a centroid set to the Gaussian fit center (same as Figure 2c). **b**, The residuals of the best radiative transfer model with a shifted centroid from minimizing the *rms* of the data minus model residuals. The asymmetry to the north-west portion of the disk is clear in both cases.

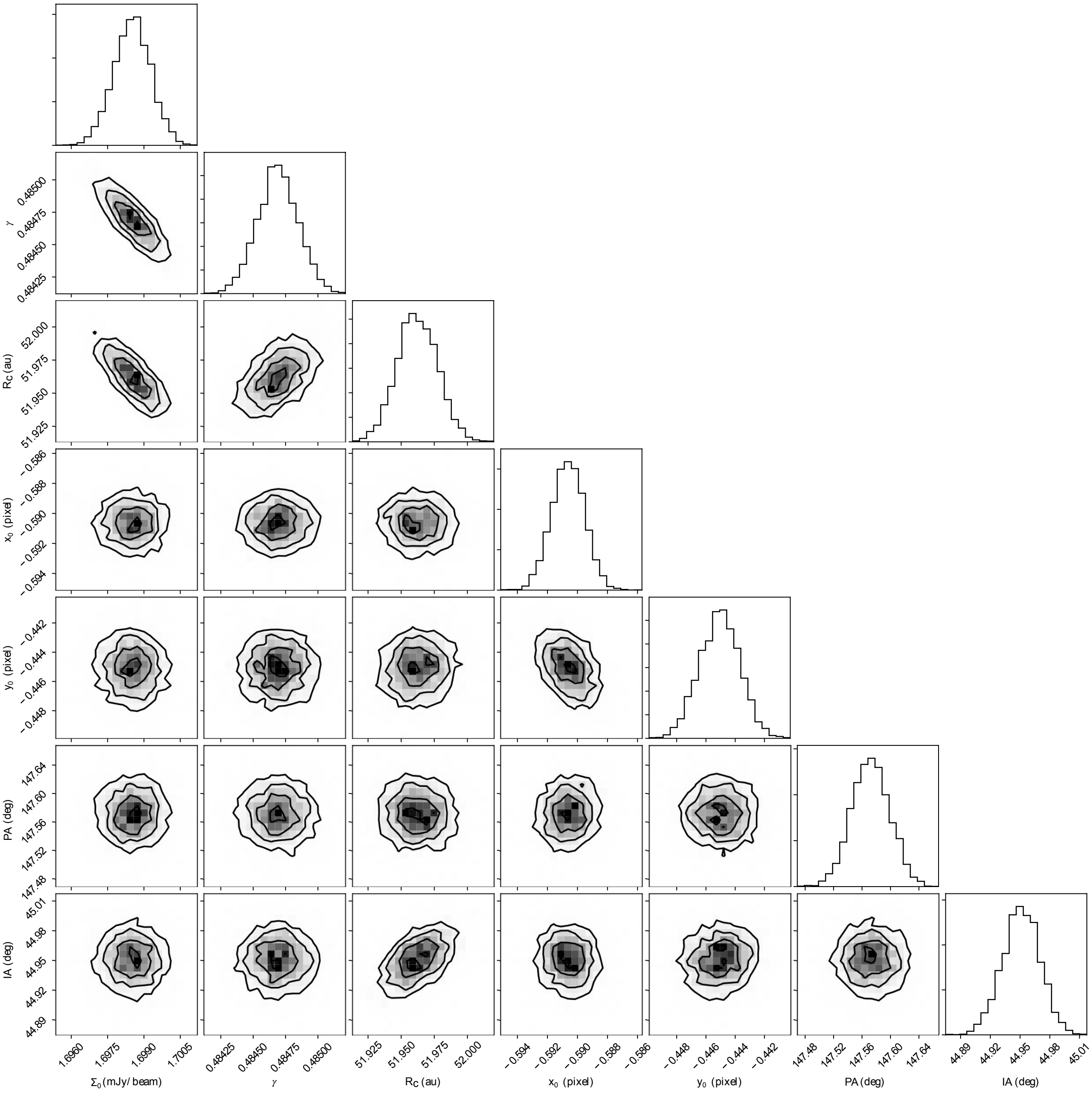

**Extended Data Figure 2 | Posterior distributions of the analytic disk model parameters.** From top to bottom, the parameters are: normalization factor $\Sigma_c$, characteristic radius $R_c$, surface density gradient $\gamma$, the coordinates of the centroid of the fit $x_0$ and $y_0$, disk position angle *PA*, and disk inclination angle *IA*.

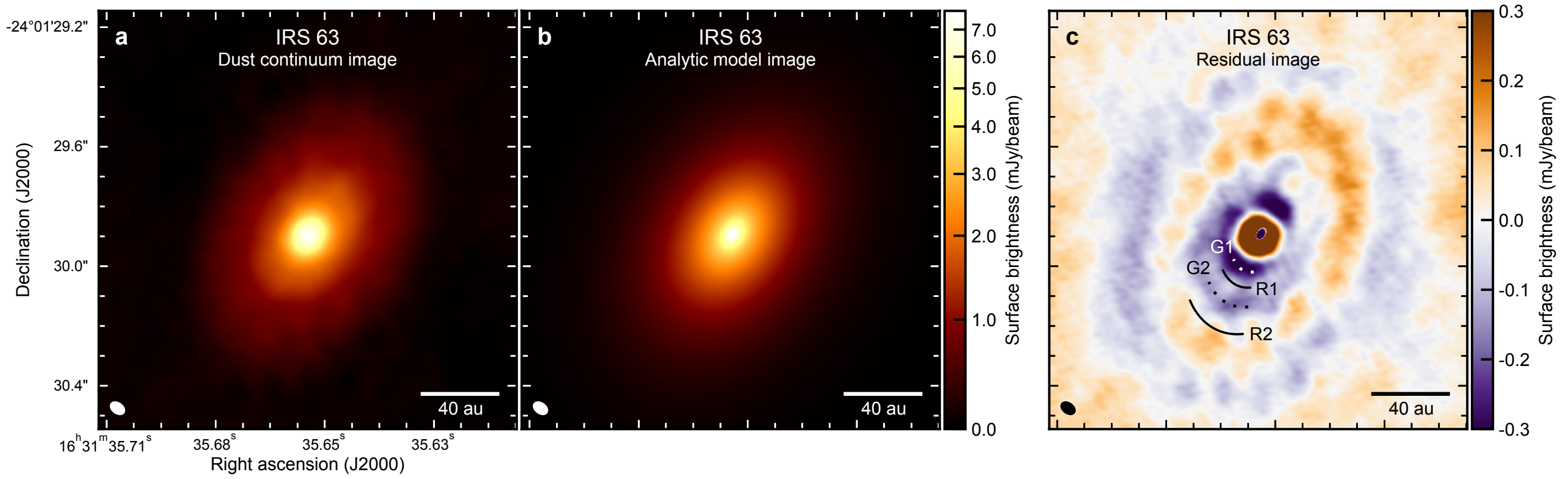

**Extended Data Figure 3 | A geometrically flat analytic model shows asymmetry in the residuals.** **a**, The original dust continuum data (same as Figure 2a). **b**, The image of the flat analytic model, generated from the median parameter values resulting from the MCMC fit. **c**, The residuals of the observed data minus the analytic model. Again, here the annular substructures and the asymmetry with excess emission to the north-west are seen.

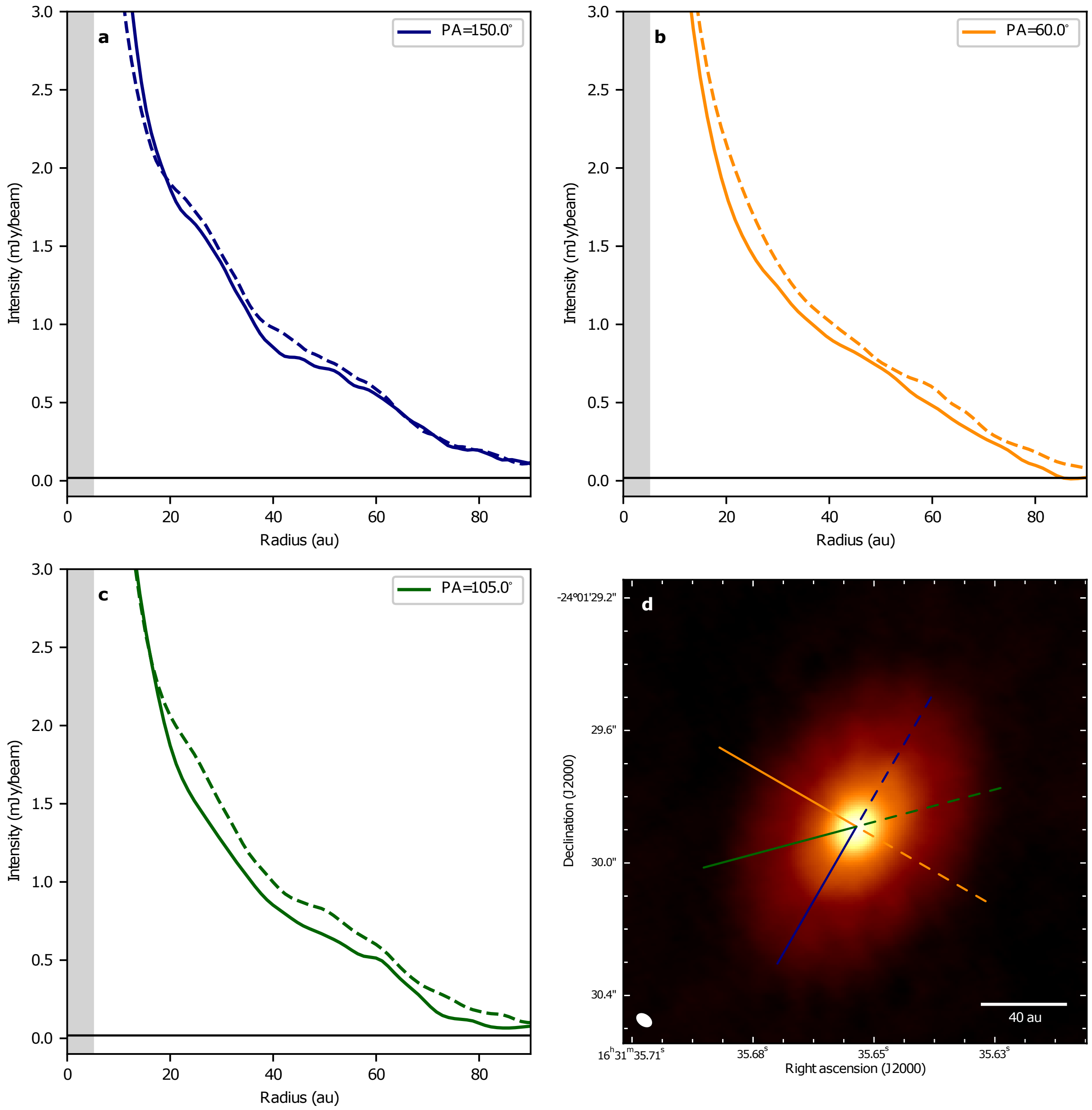

**Extended Data Figure 4 | Position-intensity cuts of the dust continuum image centered at the peak intensity position.** The centroid is (α(2000),δ(2000)) = 16 h 31 m 35.6577 s, -24° 01′ 29.897″. The cuts are taken at position angles of (**a**) 150.0° or along the minor axis, (**b**) 60.0° or along the minor axis, and (**c**) 105.0° or along a cut from the south-east to north-west. In (**d**) the orientation of these cuts across the disk are respectively shown in blue, orange, and green. The radii in **a**, **b**, and **c** are deprojected assuming a disk inclination angle of 45.0° and are measured starting from the centroid outwards. The black horizontal lines show the 1σ noise level of 18.4 μJy/beam. We plot the radial intensities along either side of the centroid for each cut (solid and dashed lines). The 105.0° cut in **c**, along the south-east to north-west direction, shows the highest difference in intensities on either side of the centroid, and is consistent with the asymmetric and excess emission to the north-west previously seen in the residuals from the disk models.

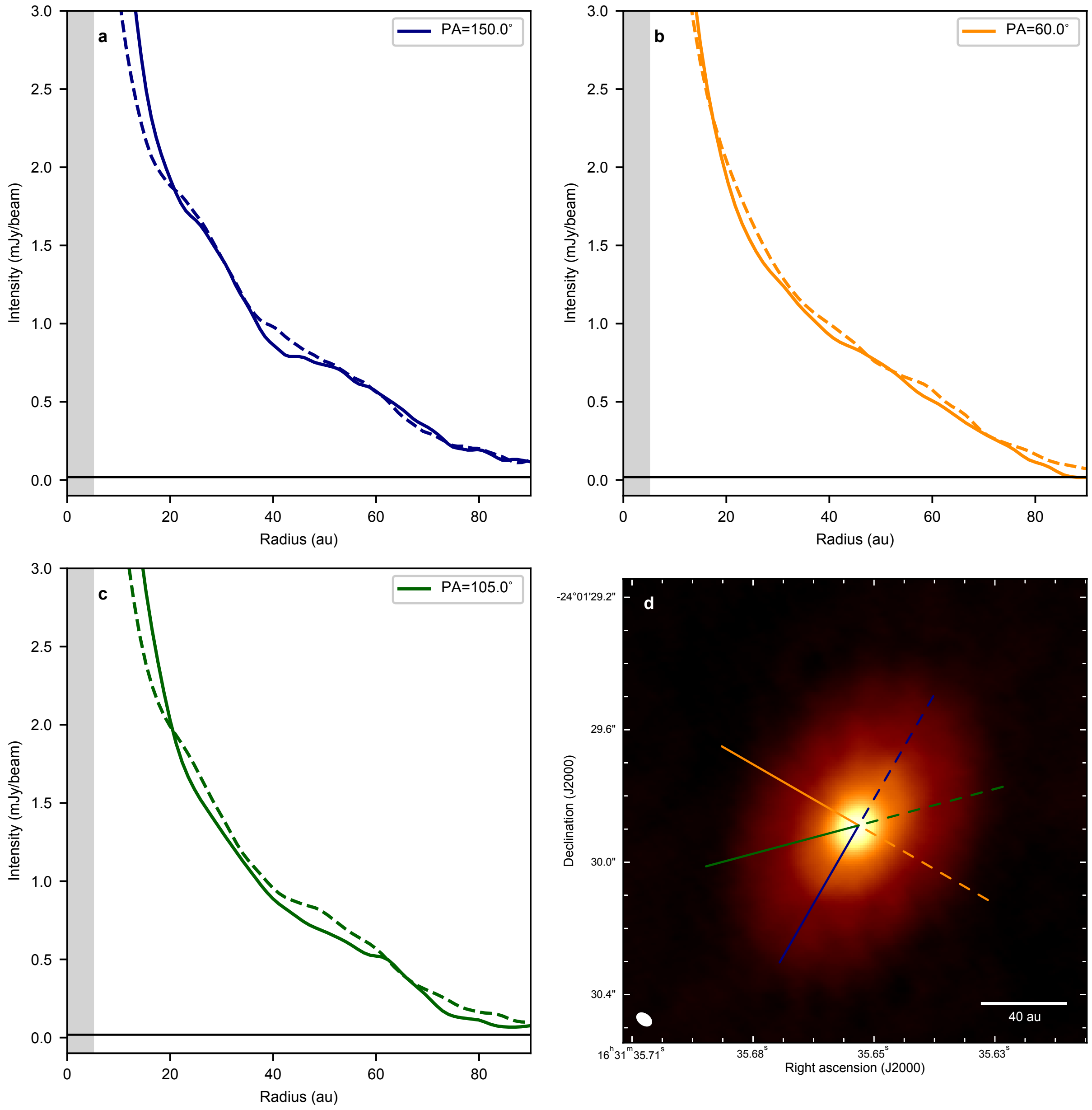

**Extended Data Figure 5 | Position-intensity cuts of the dust continuum image centered at the Gaussian fit position.** Same as Extended Data Fig. 4, but with a centroid of (α(2000),δ(2000)) = 16 h 31 m 35.6572 s, -24° 01′ 29.896″.

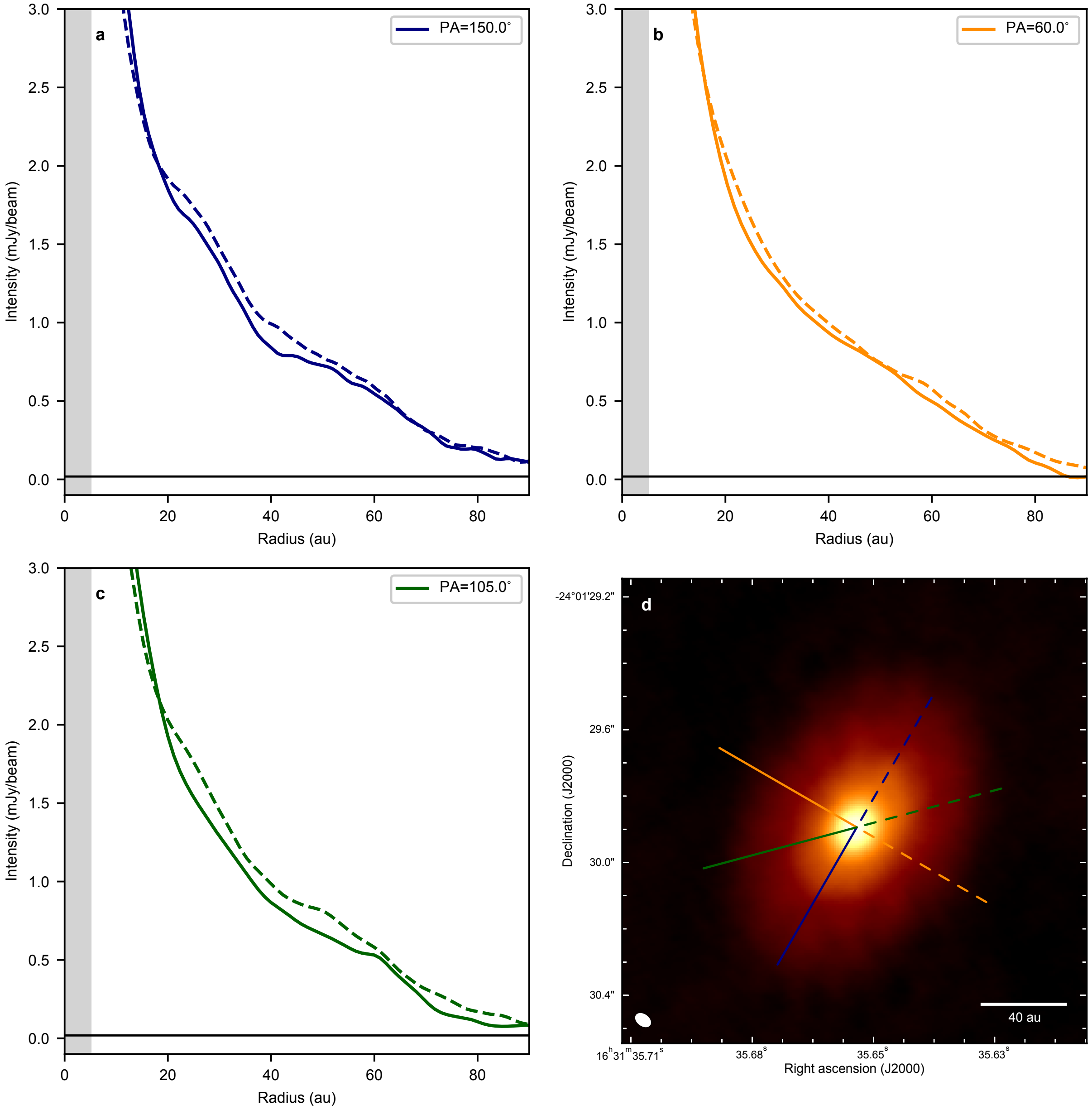

**Extended Data Figure 6 | Position-intensity cuts of the dust continuum image centered at the center determined from the MCMC fit to the analytic model.** Same as Extended Data Fig. 4, but with a centroid of (α(2000),δ(2000)) = 16 h 31 m 35.6575 s, -24° 01′ 29.901″.

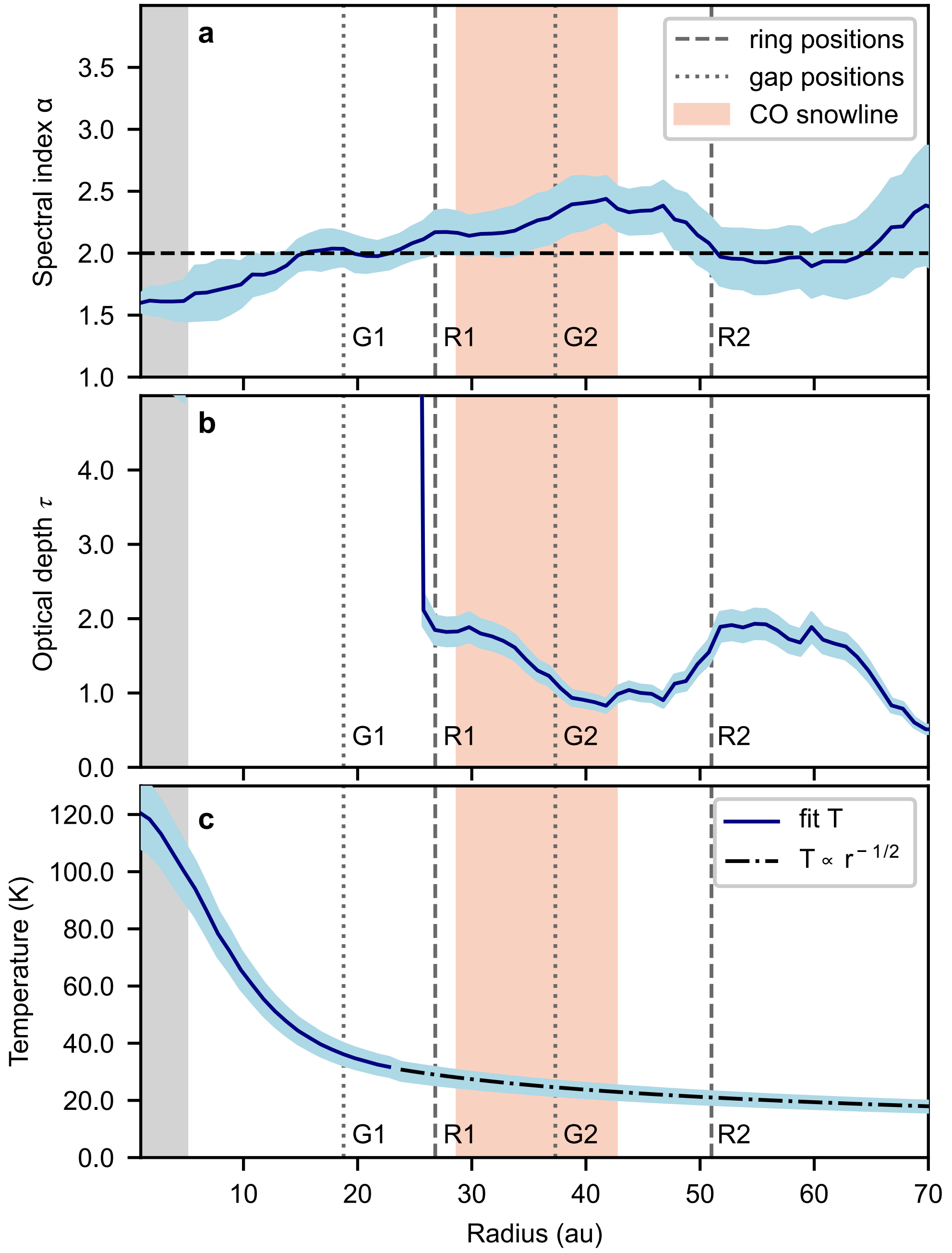

**Extended Data Figure 7 | Radial properties derived from observations.** The radial profiles of the in-band spectral index (α, **a**), the optical depth (τ, **b**), and the temperature (T, **c**) derived from the observational data. The gray shaded area represents the resolution of the observations, and the vertical grey dashed and dotted lines show the positions of the bright annular substructures R1 and R2 and dark annular substructures G1 and G2. For the spectral index the light blue ribbon represents the local standard deviation of each bin, and the horizontal dashed grey line shows the $\alpha = 2$ blackbody Rayleigh-Jeans limit. Optical depth is determined from a graybody fit to the intensity profile of the two in-band wavelengths. The temperature profile is determined from the graybody fit at radii where $\tau > 3$ (solid blue line), and extrapolated to larger radii as an inverse-square-root law (dot-dash black line). For optical depth and temperature, the light blue ribbons reflect the 10% amplitude uncertainty of the observations. The orange shaded region shows the location of the CO snowline based on dust temperature, reflecting the range of gas densities used to compute the condensation temperature.